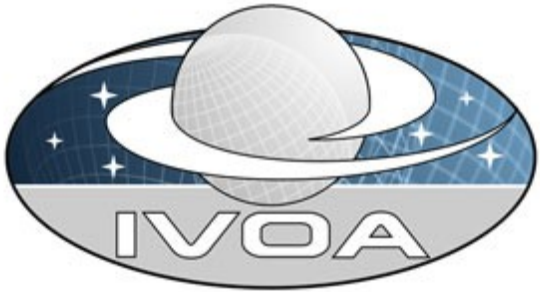

*I*nternational

*V*irtual

*O*bservatory

*A*lliance

# IVOA Single-Sign-On Profile: Authentication Mechanisms

# Version 1.01

## *IVOA Recommendation 2008 January 24*

**This version:**

1.01-2008124

**Latest version:**

http://www.ivoa.net/Documents/latest/SSOAuthMech.html

**Previous version(s):**

Proposed Recommendation:

1.01-20070906

1.0-20070621

Working Draft:

1.0-20060519

0.3 http://www.ivoa.net/internal/IVOA/IvoaGridAndWebServices/ivoa-auth-mech-0.3.doc

0.2 http://www.ivoa.net/internal/IVOA/IvoaGridAndWebServices/ivoa-auth-mech-0.2.doc

0.1 2005-06-04

**Authors:**

Grid and Web Services Working Group

Guy Rixon (editor)

## Abstract

Approved client-server authentication mechanisms are described for the IVOA single-sign-on profile: digital signatures (for SOAP services); TLS with passwords (for user sign-on points); TLS with client certificates (for everything else). Normative rules are given for the implementation of these mechanisms, mainly by reference to pre-existing standards.

## Status of This Document

This document has been produced by the IVOA Grid and Web Services Working Group.

It has been reviewed by IVOA Members and other interested parties, and has been endorsed by the IVOA Executive Committee as an IVOA Recommendation. It is a stable document and may be used as reference material or cited as a normative reference from another document. IVOA's role in making the Recommendation is to draw attention to the specification and to promote its widespread deployment. This enhances the functionality and interoperability inside the Astronomical Community.

*A list of* [current IVOA Recommendations and other technical documents](#) *can be found at http://www.ivoa.net/Documents/.*

## Acknowledgements

This document derives from discussions among the Grid and Web Services working-group of IVOA. It is particularly informed by prototypes built by Matthew Graham (Caltech/US-NVO), Paul Harrison (ESO/EuroVO), David Morris (Cambridge/AstroGrid) and Raymond Plante (NCSA/US-NVO). The prior art for the use of proxy certificates comes from the Globus Alliance.

This document has been developed with support from the National Science Foundation's Information Technology Research Program under Cooperative Agreement AST0122449 with The Johns Hopkins University, from the UK Particle Physics and Astronomy Research Council (PPARC) and from the Europena Commission's Sixth Framework Program via the Optical Infrared Coordination Network (OPTICON).

## Definitions

The **Virtual Observatory (VO)** is general term for a collection of federated resources that can be used to conduct astronomical research, education, and outreach. The **International Virtual Observatory Alliance (IVOA)** is a global collaboration of separately funded projects to develop standards and infrastructure that enable VO applications. The **International Virtual Observatory (IVO)** application is an application that takes advantage of IVOA standards and infrastructure to provide some VO service.

# Contents

# 1 Introduction

IVOA's single-sign-on architecture is a system in which users assign cryptographic credentials to user agents so that the agents may act with the user's identity and access rights. This standard describes how agents use those credentials to authenticate the user's identity in requests to services.

This document is essentially a *profile* against existing security standards; that is, it describes how an existing standard should be applied in an IVO application to support single sign-on capabilities in the IVO. In the following sections, we make specific references to details spelled out in these standards. For the purposes of validating against this standard, those referenced documents should be consulted for a full explanation of those details. Unfortunately, a reader that is unfamiliar with these external standards might find this specification confusing. To alleviate this problem, each major section is concluded by a Commentary subsection that provides some explanations of the detailed terms and concepts being referred to. The Commentary subsection may also provide recommended scenarios for how this specification might actually be realized. Note that the statements in the Commentary subsections are non-normative and should not be considered part of precise specification; nevertheless, they are indicative of the intended spirit of this document.

## 2  Scope of this standard

### 2.1  Requirements

When a service is registered in an IVO registry, that service's resource document may include metadata expressing conformance to one or more of the authentication mechanisms approved in the IVOA SSO profile. Such a service must implement those mechanisms as described in this document, and clients of the service must participate in the mechanism when calling the service.

The registration of the service interface shall contain an XML element of type *SecurityMethod* as specified in the XML schema for VOResource [VOResource]. The value of this element distinguished the authentication mechanism using the values stated in the sections below.

Services registered without the metadata alluded to above need not support any authentication mechanism. If they do require authentication, they may use either the IVOA-standard mechanisms or others that are not IVOA standards.

### 2.2  Commentary

The IVOA SSO profile allows the development of a "realm" of interoperable services and clients. Service providers opt in to this realm by implementing this current standard and by registering accordingly in the IVO registry. This allows clients to discover a secured service through the registry and to be able to use it without being customized for the details of the specific service.

Parts of the IVO that are not intended to be widely interoperable need not opt in to the SSO realm. In particular, "private" services, accessed by web browsers and protected by passwords, are allowed. However, these private services should be reworked to follow the IVOA standard if they are later promoted to a wider audience.

An example of a registration for a secured interface follows.

```
<interface xmlns:vs:="ivo://www.ivoa.net/xml/VODataService/v1.0"
           xsi:type="vs:ParamHTTP">
  <accessURL>http://some.where/some/thing</accessURL>
  <securityMethod>ivo://ivoa.net/sso/tls-with-certificate</securityMethod>
</interface>
```

## 3  Approved authentication mechanisms

### 3.1  Requirements

The following authentication mechanisms are approved for use in the SSO profile.

- No authentication required.
- Digital signature of messages.

- Transport Layer Security (TLS) with client certificates.
- Transport Layer Security (TLS) with passwords.

The mechanism is associated with the interface provided by the service and registered in the IVO registry.

- Services that are registered with a VO registry as having a *WebService* type interface [VOResource] shall support digital signature, or shall support TLS with client certificates or shall require no authentication.
- Interfaces by which a user logs in to the SSO system shall support either TLS with client certificates, or TLS with passwords, or both.
- All other interfaces shall support TLS with client certificates or shall require no authentication.

## 3.2 Commentary

Services with interface type *WebService* are SOAP services.

Digital signatures on messages are preferred as (a) they protect the message through a chain of handlers and (b) they avoid the need, in conventional TLS, to certify the service. However, most message formats have nowhere to convey a signature. The envelope structure in SOAP, consisting of a message body and separate header, is ideally suited to carrying signatures.

The digital-signature and TLS-with-client-certificate mechanisms allow the service to verify that the client holds the private key matching a certificate transmitted with the message. Authentication succeeds if the service trusts the issuer of the certificate. That trust is determined by reference to a set of certificates for trusted certificate authorities (CAs) configured into the service by the service provider.

# 4 Details of TLS

## 4.1 Requirements

Services using Transport Layer Security (TLS) shall do so according to the TLS v1.0 standard [RFC2246].

## 4.2 Commentary

TLS is derived from the earlier standard known as Secure Sockets Layer of which versions 2.0 and 3.0 are in use at the time of writing. TLS v1.0 is based on SSL v3.0; "*the differences between [TLS v1.0] and SSL 3.0 are not dramatic, but they are significant enough that TLS 1.0 and SSL 3.0 do not interoperate (although TLS 1.0 does incorporate a mechanism by which a TLS implementation can back down to SSL 3.0)*" [RFC2246]. SSL v2.0 has known flaws that compromise its security.

# 5 Details of TLS-with-client-certificate

## 5.1 Requirements

Certificates shall be transmitted and checked according to the TLS v1.0 standard [RFC2246].

Services implementing TLS must support certificate chains including proxy certificates according to RFC3820 [RFC3820].

Interfaces using this mechanism shall be be registered with the security method *ivo://ivoa.net/sso#tls-with-client-certificate*.

# 6 Details of TLS-with-password

## 6.1 Requirements

The user-name and password shall be passed in the message protected by the TLS mechanism, not as part of the mechanism itself. In particular, "HTTP basic authentication" shall not be used.

Interfaces using this mechanism shall be be registered with the security method *ivo://ivoa.net/sso#tls-with-password*.

## 6.2 Commentary

"HTTP basic authentication" passes the user-name and password in the HTTP headers, assuming that the credentials are not a natural part of the message body. This standard applies the TLS-with-Password mechanism only to the special case of logging in to the SSO realm. Hence, the user name and password are logically part of the message body, not the message header.

# 7 Details of digital signature

## 7.1 Requirements

SOAP clients shall sign digitally the entire body of a SOAP request message and shall send the signature to the SOAP service as part of the message. The signature shall be encoded in the SOAP header according to the rules of the WS-Security standard [WS-Security].

The client shall sign messages using a certificate chain as described in section 8 of this document. This chain shall be included in the SOAP header as an element of type *BinarySecurityToken* (as defined in [WS-Security], section 6.3) The signature in the SOAP header shall refer to this *BinarySecurityToken* by including a suitable *SecurityTokenReference* (as defined in [WS-Securtity], section 7.1) in the signature.

The service shall authenticate the sender of a signed message in a two-step process:

1. validate the signature on the message body;

2. reconstitute and validate a chain of certificates leading from the certificate associated with the message signature back to a trust anchor (see below) trusted implicitly by the service.

The rules for validating the signature are given in the WS-Security standard and in standards referred by WS-Security. The rules for validating the chain of trust are described in section 8 of this document.

Interfaces using this mechanism shall be be registered with the security method *ivo://ivoa.net/sso#soap-digital-signature*.

## 7.2 Commentary

A trust anchor is a self-signed certificate that the application trusts as being authentic. This is usually because the certificate was obtained by the application via a different channel (usually explicitly by the application's administrator) than the channel connecting the service and the client. See section 8.2 for more discussion of the role of a trust anchor.

WS-Security, in respect of digital signatures, is mainly a set of rules for organizing the XML elements defined by IETF digital-signature standard [RFC3275]. The IETF standard is very general; WS-Security "blesses" a few of the possible patterns of usage. This profile in turn "blesses" a few of the possible uses of WS-Security.

There are two extant versions of WS-Security, v1.0 and v1.1. These versions specify the same XML for a signed message, so the question of which version to support is not important.

# 8 Certificate chains

## 8.1 Requirements

The certificate chain sent by a client to a service during authentication shall contain only end-entity certificates [RFC2459] and proxy certificates [RFC3820]; both are forms of the X.509v3 certificate format. The transmitted chain shall not include any self-signed certificates. The identity which the client wishes authenticated must be the subject of the first end-entity certificate in the chain.

For the purpose of establishing trust, the service shall be configured with one or more chains of certificates each of which ends with a trust anchor (see section 8.2). These configured chains must contain only end-entity certificates.

To establish trust in the identity with which a message is signed, a service shall concatenate the chain transmitted by the client with one of the configured chains, forming a single chain from the certificate used in the message signature back to the trust anchor. The service shall then validate the relationship between each two links in the chain according to the IETF rules [RFC3820].

A certificate may contain a limit on the length of the chain of proxy certificates that may be derived from it. The validation of the certificate chain shall respect this limit if it is present.

If the chain is validated successfully, then the service shall deem the chain to authenticate the identity that is the subject of the first end-entity certificate in the chain.

## 8.2 Commentary

In a chain of certificates, each certificate is signed by the certificate that comes after it, or is self-signed. Self-signed certificates trusted by the service are called "trust anchors"; they are generally certificates issued by certificate authorities (CAs). End-entity certificates are issued on behalf of users (humans or automata) by CAs and are often "permanent" – that is, they have a long lifetime. Proxy certificates are usually short-term credentials (valid typically for a few hours or days).

Proxy certificates are often used to solve two problems. The first is that a user may authenticate herself using a long lived EEC; however, it would not be safe to allow an application to have long-term use of that certificate, particularly if that application is remote. To solve this, a short-lived proxy certificate can be provided to the application to limit its privileges. Note that in some authentication models, the user can authenticate with a delegation service which can provide the application with a certificate. This can be either a short-lived proxy or a short-lived EEC; either will solve the problem.

The other problem is that one service requiring authentication will need to call another service requiring authentication. Solving this problem in the IVOA SSO framework requires the use of proxy certificates. Below is a scenario for how this can be accomplished.

A user agent signs on to the SSO system (the details of the sign-on process are not specified in this document) in a user's name and receives a chain of two certificates: a proxy and an end-entity certificate. The user-agent holds the private key matching the proxy certificate but not that matching the end-entity certificate. The agent signs messages with the proxy and includes in the message the end-entity certificate in order to complete the chain back to the CA certificate that is the trust anchor.

When service B receives an authenticated request from agent A (signed with a proxy certificate), and when makes an authenticated call to service C while satisfying A's request, then B derives from A's proxy a new proxy to which B holds the private key (the protocol for deriving this proxy is specified outside this document). This new proxy is signed using A's original proxy. Therefore, C receives a chain of two proxy certificates and one end-entity certificate. If C then calls service D, C sends a chain of three proxies and one end-entity certificate.

Some of the chain-validation rules from RFC3820 can be summarized as follows (this is an incomplete description of the rules and is non-normative for the IVOA SSO profile).

- A proxy certificate may be followed by an end-entity certificate or another proxy certificate.

- An end-entity certificate may not be followed by a proxy certificate.
- There must be exactly one self-signed certificate and it must be the last in the chain: this is the trust anchor.
- Any proxy certificate must carry the *ProxyCertInfo* extension (and must therefore be in X.509v3 format).
- Any end-entity certificate that follows another end-entity certificate must carry the basic-constraints extension with the CA field set to true (and must therefore be in X.509v3 format).
- There must be at most one end-entity certificate that does not have the CA field in the basic-constraints extension set to true. This must be the first end-entity certificate in the chain.

# 9  Changes since previous version of this document

- All references to "identity certificate" (a Globus term) are changed to "end-entity certificate" (an IETF term defined in RFC3820).
- The form of registration for secured interfaces is specified.